\documentstyle[twocolumn,epsfig,prb,aps,amsmath,amssymb]{revtex}

\begin{document}

\twocolumn[\hsize\textwidth\columnwidth\hsize\csname
@twocolumnfalse\endcsname

\title{Quantum Monte Carlo simulations of the t-Jz model with
stripes on the square lattice}

\author{Jos\'e A.~Riera}
\address{
Instituto de F\'{\i}sica Rosario, Consejo Nacional de 
Investigaciones 
Cient\'{\i}ficas y T\'ecnicas, y Departamento de F\'{\i}sica,\\
Universidad Nacional de Rosario, Avenida Pellegrini 250, 
2000-Rosario, Argentina}
\date{\today}
\maketitle
\begin{abstract}
Finite temperature quantum Monte Carlo simulations are performed
on the anisotropic $t-J$ model and in particular on its Ising limit.
Straight site-centered stripes are imposed by an on-site potential
representing external mechanisms of stripe formation.
In this model, we show that, even though charge inhomogeneity exists
at a high temperature, the anti-phase ordering of the spin domains
between stripes occurs at a much lower temperature. The magnetic
correlations at this spin ordering crossover are analyzed.
The stripes show metallicity,
with absence of hole attraction. Comparison between this model
and others that have been proposed to explain or describe
stripes, as well as possible relations with experimental features
on underdoped cuprates are discussed.
\end{abstract}

\smallskip
\noindent PACS: 71.10.Fd, 74.80.-g, 74.72.-h

\vskip2pc]

The nature of the underdoped region and of the pairing mechanism
in high-T$_c$ cuprates is still a matter of strong controversies.
From the experimental point of view, the relationship between the
pseudogap (PG) and the superconducting gap (SG) is still under
intense discussion.
Older data suggested a pseudogap smoothly connecting with the
superconducting gap\cite{puchkov} while some more recent
studies\cite{tallon,PGnotS} emphasize a different origin and
behavior of the PG and the SG, the PG ending inside the
superconducting phase at a ``quantum critical point".\cite{qcp} 
From the theoretical point of view,
the stripe scenario, which is an almost unavoidable consequence of
neutron experiments results on
La$_{2-x}$Sr$_x$CuO$_4$\cite{tranquada,ichikawa} and perhaps on 
YBa$_2$Cu$_3$O$_{6+\delta}$\cite{mook}, offers a natural
explanation for PG.
Consistently with the experimental controversy about the nature
of PG, there are widely diverging views about the origin of stripes
and its relationship with superconductivity.
In the theory by Emery, Kivelson and
coworkers\cite{emery}, the pseudogap is identified with the SG
and stripes are a key ingredient to explain
superconductivity in the cuprates. In another view,\cite{white}
stripes are regarded as competing with a uniform gas of hole pairs
and hence with superconductivity. In yet another approaches, the
stripes exclude hole pairing.\cite{martins} There are even more
important differences regarding the origin of stripes.
In Ref.~\onlinecite{emery}, charge inhomogeneity appears as
a combined effect of phase separation and long-range Coulomb
repulsion and the stripe phase can be thought as a Wigner crystal.
According to White and Scalapino\cite{white}, stripes are already
present in the simple $t-J$ model at physical values of the
parameters. In Ref.~\onlinecite{martins}, the driving mechanism
is the formation of strong singlets across a hole. In this sense,
the stripes may be regarded as domain walls.\cite{zaanen} On the
other hand, in some other views, the stripes are not inherent to
two-dimensional (2D) extended $t-J$ models but are due to,
for example, electron-phonon coupling.\cite{castroneto,petrov}
Following these alternative views, in this paper we formulate a model
assuming that the origin of stripes in the cuprates is non-intrinsic
to the 2D electronic correlations and our main goal is to examine
the physically relevant properties of such model.

It is also well known the difficulty of studying microscopic models
of correlated electrons either by analytical or numerical techniques.
The already
mentioned work on variants of the $t-J$ model by using the DMRG
method\cite{white,martins} is affected by limitations of this
method, particularly the open boundary conditions adopted. The
widely used Lanczos diagonalization could not deal with clusters
large enough to reproduce the charge inhomogeneities. In the
present study, we use the conventional finite temperature quantum
Monte Carlo (QMC) method (world-line algorithm)\cite{reger}
which allows the study of reasonable large clusters with fully
periodic boundary conditions.
As it is well-known, QMC simulations of fermionic models are
affected by strong ``minus sign problem"\cite{minus} that virtually
makes impossible these kind of studies at very low temperatures.
However, as we discuss below, in the model we consider the ``minus
sign problem" is not as much severe as in the plain 2D $t-J$ model,
and hence we are able to look at physical features which appear
at not too low temperatures.

Let us first introduce the model here studied. In the first place, as
in many other studies on this
subject\cite{tworzydlo,kim,shibata,liu,eroles},
we impose the presence of straight site-centered stripes by an
on-site potential. This on-site potential represents the effects
of Coulomb potential due to out-of-plane ions, electron-phonon
coupling, a-b plane anisotropy or other structural details, etc.
Since in-plane long-range Coulomb repulsion can not be included in
QMC simulations, its effects can eventually also be represented by the
on-site potential. The confinement of holes to the stripes strongly
reduces the ``minus sign problem". As we show below, this problem
strongly constrains the values of the on-site potential accessible
in the simulations.  A further alleviation of this problem comes from
reducing quantum spin fluctuations.  We are thus lead
to an anisotropic $t-J$ model, and in the Ising limit to the
so-called $t-Jz$ model.\cite{tjz,tjz-strip} Again, the ``minus sign
problem" strongly constrains the off-diagonal exchange term we
can deal with. On the other hand, this Ising 
limit is not only convenient from the numerical point of
view but has an additional value, specially if one is interested
in finite temperature effects. In fact, a three-dimensional (3D) AF 
(short- or long-range) order would imply, at a mean-field level,
a staggered field acting on the 2D t-J model and in turn
would induce an enhancement of the $zz$-component of the Heisenberg
term as a second order process.

The Hamiltonian of the anisotropic $t-J$ model is:
\begin{eqnarray}
H_{tJ} =
&-& t \sum_{ \langle { i j} \rangle,\sigma }
({\tilde c}^{\dagger}_{ i\sigma}
{\tilde c}_{ j\sigma} + h.c. ) 
\nonumber  \\
&+& J \sum_{ \langle { i j} \rangle }
( \frac{\gamma}{2} (S^{+}_{i} S^{-}_{j} +
S^{-}_{i} S^{+}_{j}) +
S^{z}_{i} S^{z}_{j} -
{\frac{1}{4}} n_{i} n_{j} )
\label{ham_anis}
\end{eqnarray}
\noindent
where the notation is standard. The stripes are induced by an
effective on-site potential:\cite{castro-pot}
\begin{eqnarray}
H_{str}  = \sum_{i} {e_s}_i n_{i}
\label{stripepot}
\end{eqnarray}
\noindent
${e_s}_i=-2 e_s < 0$ ($2 e_s$) for sites on (outside) the stripe.
Then, the total Hamiltonian of our model is $H=H_{tJ}+H_{str}$.
Most of our calculations were performed on an $8 \times 8$
cluster
with fully periodic boundary conditions and with eight holes which
corresponds to a filling of $x=1/8$. The imposed stripes involves
columns separated by three-leg ladders, as in the original
picture in Ref.~\onlinecite{tranquada}.

Our simulations were done at $J/t=0.35$, a value generally accepted
to describe the cuprates, and also at $J/t=0.7$. In this second
case, we have seen essentially the same physical behavior but at a
higher temperature scale and with a milder minus sign problem. As
usual, $t$ is chosen as the unit scale of energy and temperature.
$e_s$ was varied between 0.3 and 2.0, and $\gamma=0.0$ (the Ising
limit), 0.25 and 0.5 were examined.
In the limit
$e_s \rightarrow \infty$ each stripe would be at quarter filling.
Simulations were also performed for the $12 \times 12$ cluster
with twelve holes and also two equidistant stripes.
This corresponds to a smaller density $x=1/12$ and the
spin domains in between the stripes are five-leg ladders.

\begin{figure}
\begin{center}
\epsfig{file=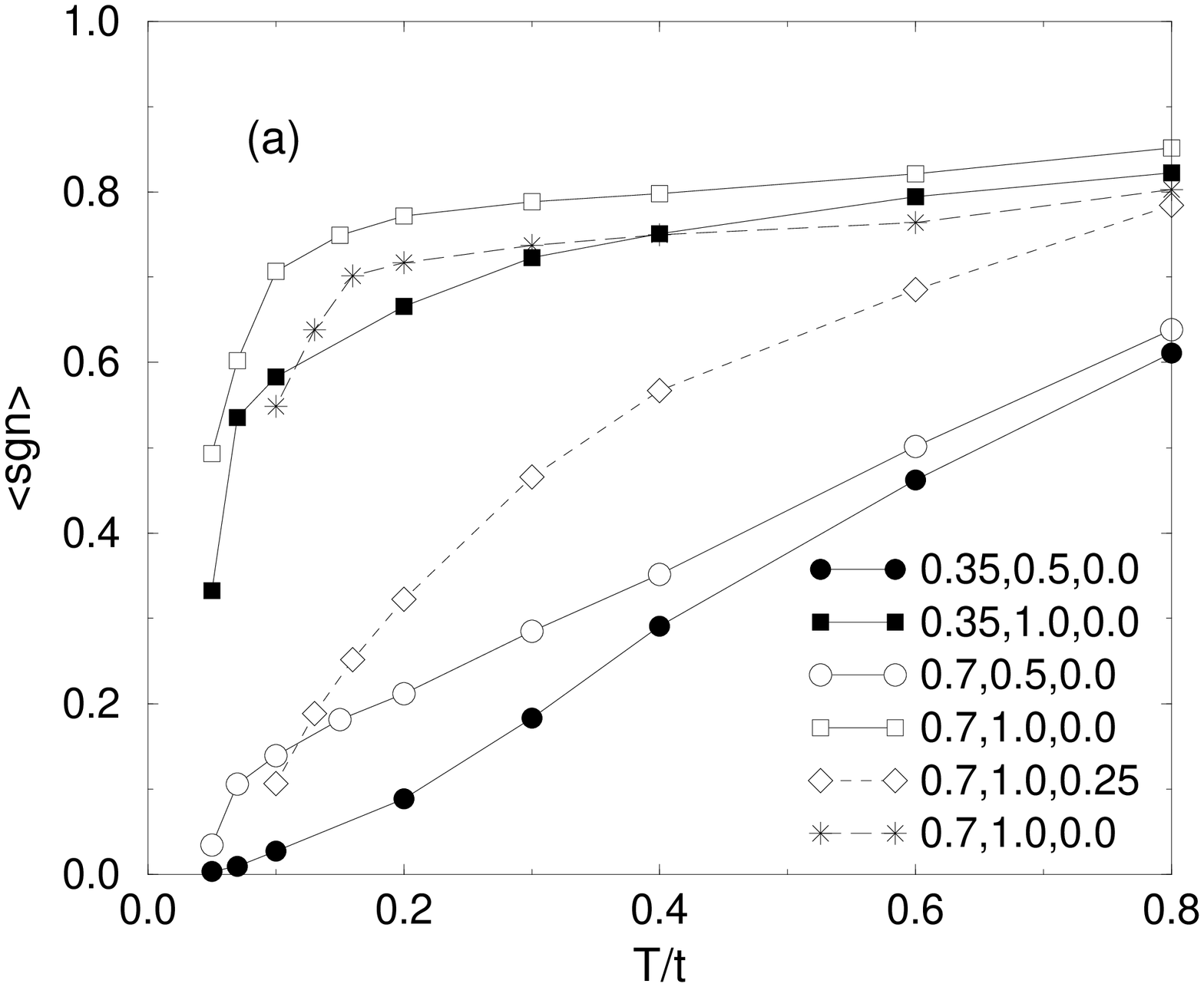,width=7.0cm,angle=-90}
\epsfig{file=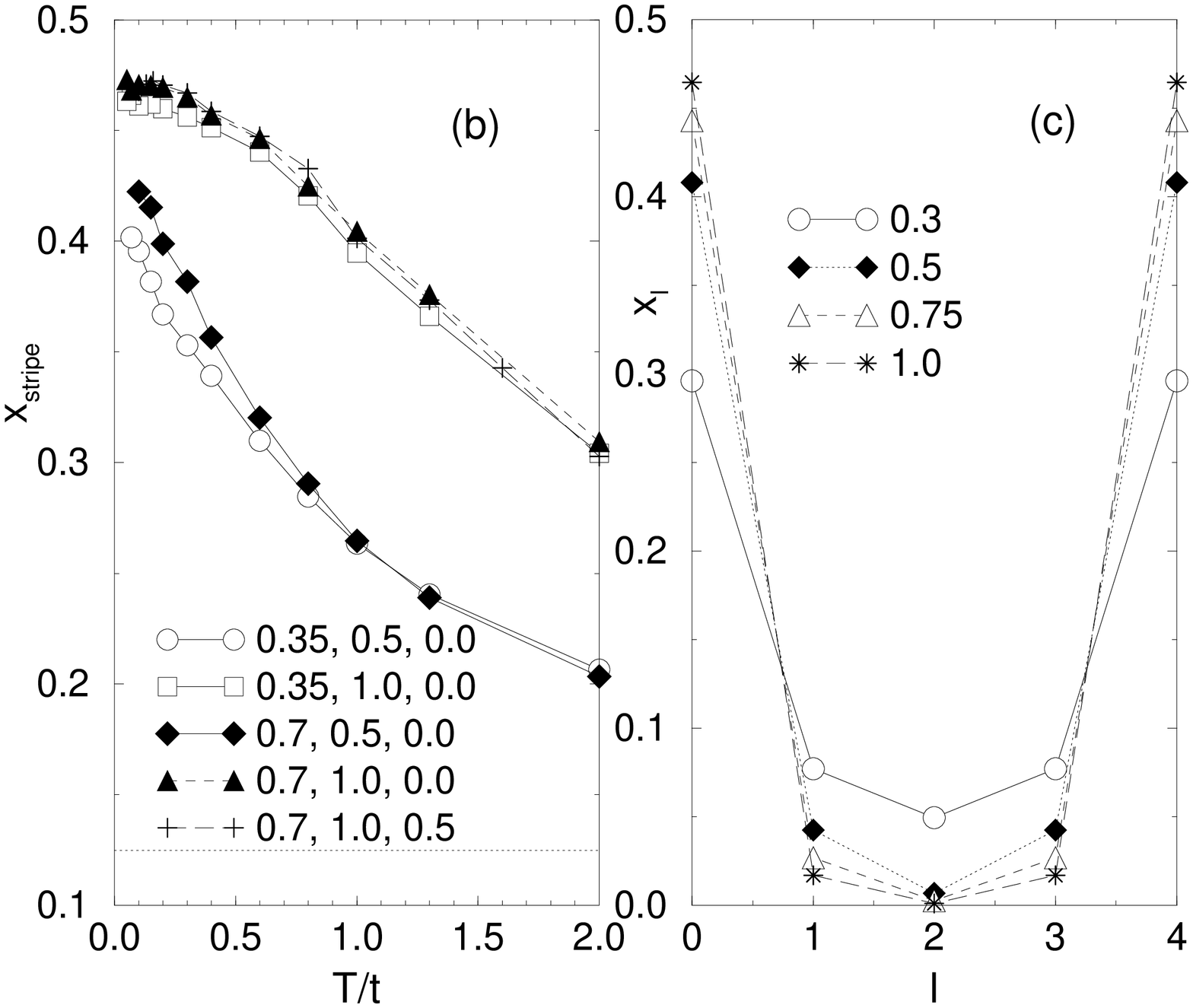,width=7.0cm,angle=-90}
\end{center}
\caption{
(a) Averaged sign in QMC simulations on the $8 \times 8$ cluster
with eight holes.
The sets ($J$,$e_s$, $\gamma$) are indicated on the plot.
Stars correspond to $12 \times 12$, twelve holes ($x=0.083$).
(b) Hole density on the stripes vs. temperature, $8 \times 8$ 
cluster. The dotted line corresponds to the uniform hole distribution
($x=1/8$).
(c) Hole density profile ($l=0$ and 4 correspond to the stripes) for
the $t-Jz$ model with $J=0.35$, $T<0.1$ and for several values of $e_s$
as indicated on the plot ($T=0.15$ for $e_s=0.3$).
}
\label{fig1}
\end{figure}

The QMC algorithm employed is a straightforward extension of the
world-line one successfully used to study the 2D Heisenberg
model.\cite{reger} Besides the cube and plaquette local moves,
we have kept the global moves that change the total $S_z$ of the
system. Most of the calculations were performed with
$\tau=\beta/M=0.083$, where $\beta=1/T$ and $M$ is the Trotter
number.
The average of the sign of $\exp{-\beta H}$, is shown in
Fig.~\ref{fig1}(a). It can be seen that $\langle sgn \rangle$
is smaller for larger $\gamma$ and for smaller $e_s$. In
few words, the more isotropic in the spin space and the more
homogeneous is the hole movement in real space, the worse becomes 
the minus sign problem.
In Fig.~\ref{fig1}(b), the hole occupancy of sites on the stripes
are shown as a function of temperature for various sets of
parameters ($J$,$e_s$, and $\gamma$).
As expected, the stripe filling increases monotonically as T
decreases. At large temperature there is already a filling of the
stripes larger than the nominal one ($x=1/8$) and it is
driven solely by the on-site potential. As the temperature goes
to zero, the stripe density saturates at a value smaller than
$1/2$ and it is apparent a small dependence on $J$, i.e., there
is a correction due to the magnetic correlations of the $t-J$
model. In Fig.~\ref{fig1}(c), the hole
density profile is shown for $J=0.35$ and several values of $e_s$
at the lowest temperatures reached. The sharpness of these
profiles could be measured by neutron scattering.\cite{note0}

\vspace{-0.2cm}

\begin{figure} 
\begin{center}
\epsfig{file=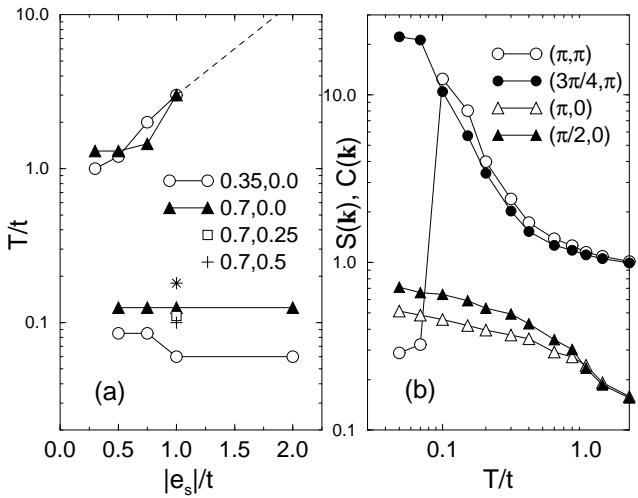,width=6.7cm,angle=-90}
\end{center}
\caption{
(a) Phase diagram in the temperature-on-site potential plane. The
  curves at high (low) T correspond to a crossover in the charge
 (magnetic) structure factor.
The sets ($J$, $\gamma$) are indicated on the plot. The
star corresponds to $12 \times 12$, $x=0.083$,
 $J=0.7$, $\gamma=0.0$.
(b) Magnetic and charge static structure factors vs. temperature
for $J/t=0.35$, $|e_s|=0.5$, $\gamma=0$}
\label{fig2}
\end{figure}

We start now to show the main features observed.\cite{monte2} 
Results of the computation of magnetic and charge static structure
factors, $S({\bf k})$ and $C({\bf k})$ (Fourier transformed
spin-spin and hole-hole correlations functions respectively) are
partially summarized in Fig.~\ref{fig2}(a). In the $8 \times 8$
cluster ($x=1/8$), at a rather large T there is a crossover in the
peak of the structure factor from $(k_x,k_y)=(\pi,0)$ (with the
$x$-axis perpendicular to the stripes direction) to $(2\delta,0)$
with $\delta=\pi/4$. This crossover, together with the behavior
shown in Fig.~\ref{fig1}(b) indicates a non-trivial behavior of the
charge ordering. One is tempted to term this crossover as the 
``charge ordering" temperature,\cite{t1star} although this concept
is somewhat arbitrary in our model.
The most important feature is that at a temperature
much lower that this crossover there is a second crossover in
the spin sector signalled by a change in the peak of the magnetic
structure factor from $(\pi,\pi)$ to $(\pi \pm \delta, \pi)$.
This peak very much resembles the one observed in
neutron scattering experiments\cite{tranquada,mook} signalling the
presence of the ``incommensurate phase" in underdoped cuprates.
Following Ref.~\onlinecite{emery}, we call $T^\ast_2$ this lower
crossover at which a ``spin ordering" occurs.
A similar behavior is observed in the $12\times 12$ lattice with
12 holes ($x=0.083$). Following the behavior observed in
underdoped cuprates, $T^\ast_2$ occurs at a temperature
higher than the one observed for similar parameters but for
$x=0.125$. The peaks of $S({\bf k})$ and $C({\bf k})$ are the same
as above except that for this smaller hole doping, $\delta=\pi/6$.
A more detailed evolution of the peaks with temperature is shown
in Fig.~\ref{fig2}(b). In particular, the crossover in the charge
sector is rather smooth. In the spin sector, the weight of the
$(\pi,\pi)$ peak is strongly reduced below the crossover, but
$S(\pi \pm \delta, \pi)$ is definitely nonzero above it.

We have detected important hysteresis effects in the crossover
region. The results shown in Fig.~\ref{fig2}(b) were obtained by
starting the simulation from a configuration with $\pi$-shifted
spin domains. If the starting configuration consists of in-phase
spin domains the crossover temperature is pushed to smaller
values. In this case, for many values of the parameters
$J$, $e_s$ and $\gamma$ (specially for large $J$ and $e_s$) we 
could not detect that crossover to the lowest temperature
attainable. In most cases, starting from a randomized spin
configuration, behaviors like those of Fig.~\ref{fig2} are
recovered but we cannot rule out the possibility
of in-phase domains being more stable than or degenerate with
anti-phase domains\cite{mook,in-phase} at low temperatures
and for very large $J$ and $e_s$.

\begin{figure}
\begin{center}
\epsfig{file=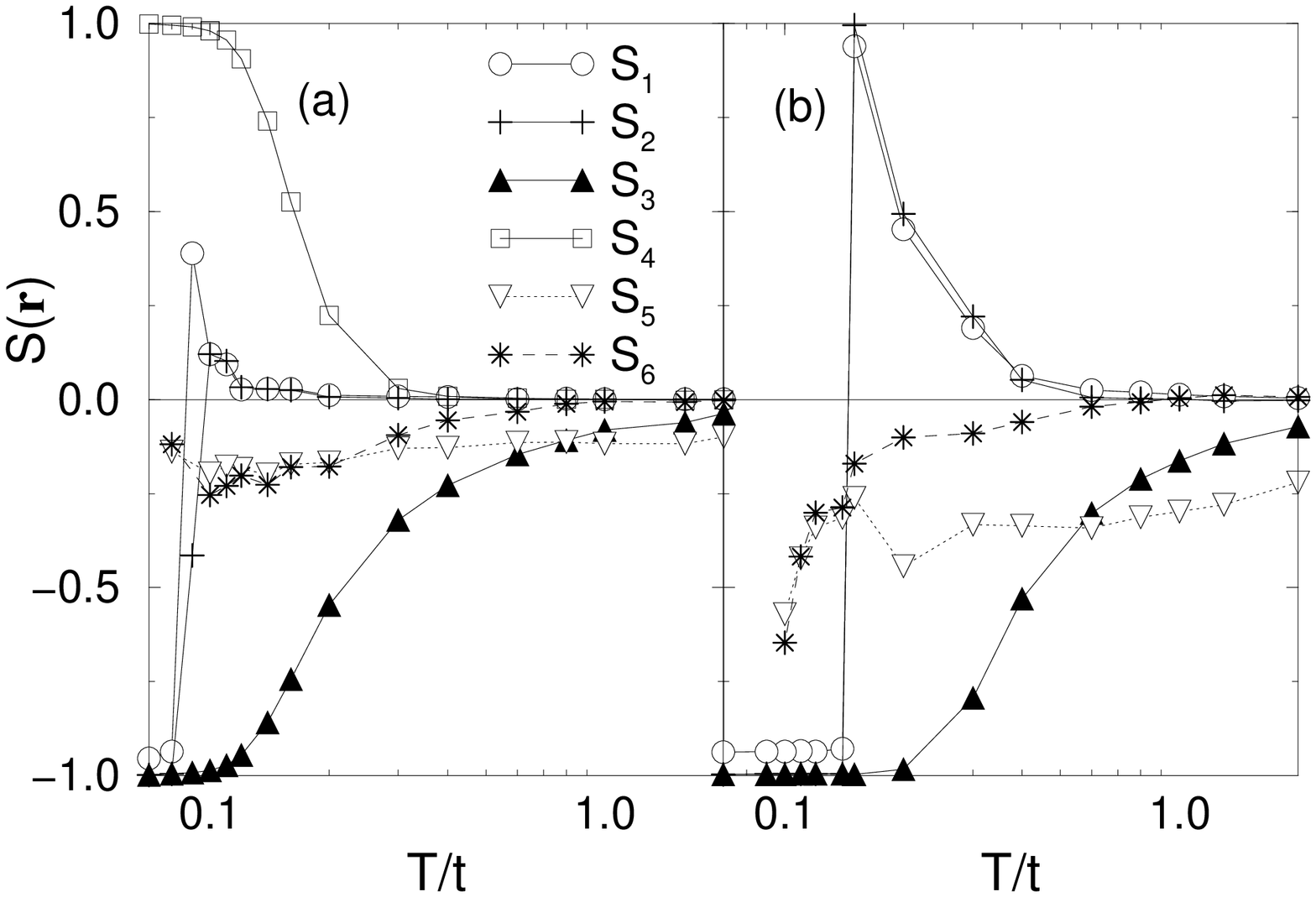,width=5.9cm,angle=-90}
\epsfig{file=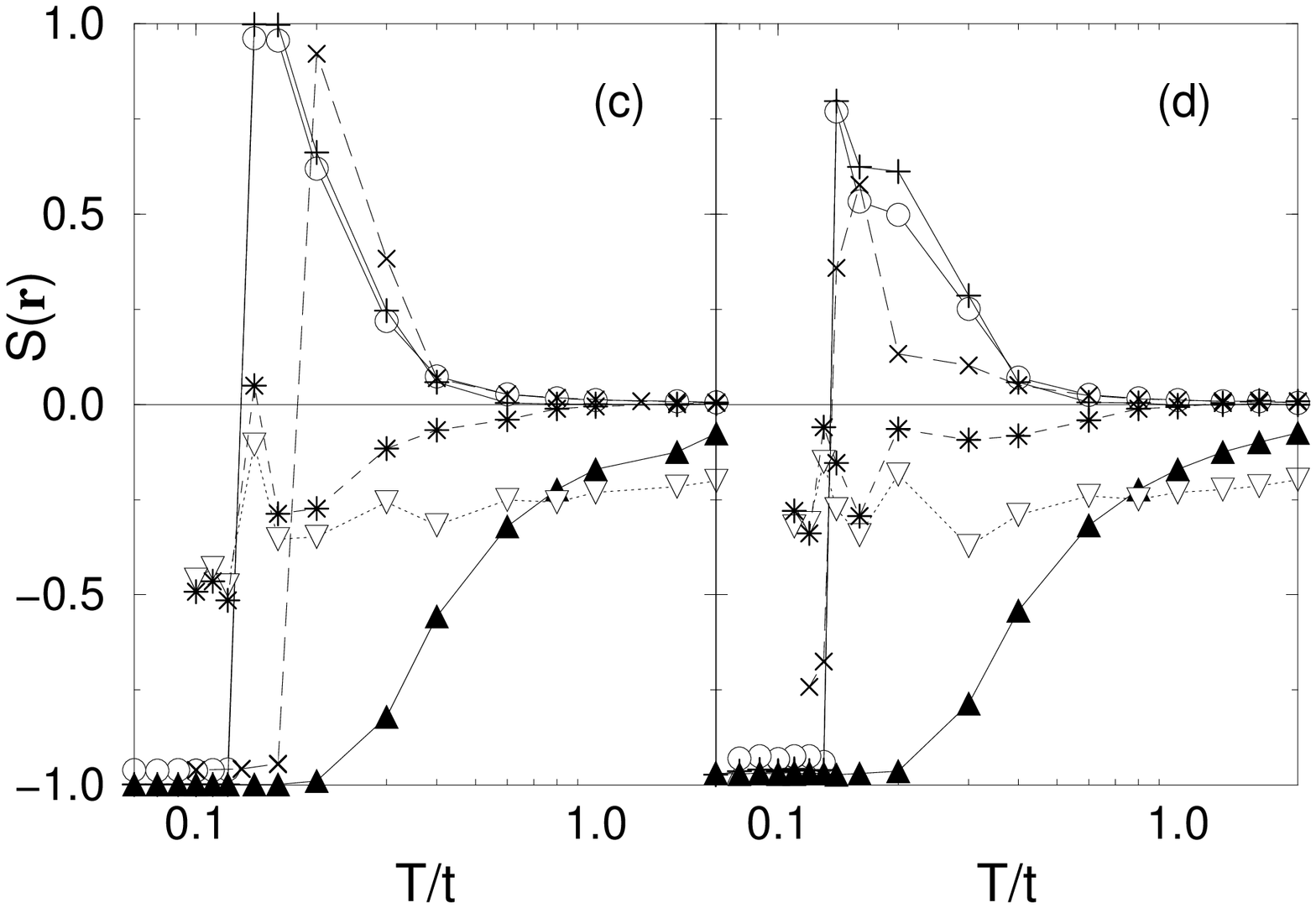,width=5.9cm,angle=-90}
\end{center}
\caption{Spin-spin correlations (defined in the text) vs.
 temperature on the $8\times 8$ cluster, for (a) $J/t=0.35$,
 $e_s=1.0$, $\gamma =0$, (b) $J/t=0.7$, $e_s=0.75$, $\gamma =0$, (c)
 $J/t=0.7$, $e_s=1.0$, $\gamma =0.0$ (crosses: $S_1$, $12\times 12$
  cluster), and (d) $J/t=0.7$, $e_s=1.0$, $\gamma =0.25$. (crosses:
 $S_1$, $\gamma =0.5$). Correlations $S_5$ and $S_6$ have been
multiplied by 5.}
\label{fig3}
\end{figure}

We can analyze the crossover in the spin sector at a more
microscopic level by looking at the real-space spin-spin correlations
that experience the most important changes at this crossover.
These correlations are between sites two lattice spacings apart
in the same row across a stripe (labelled $S_1$), and between
sites belonging to the center leg of two consecutive three-leg 
ladders in the same row ($S_2$). For completeness
we have also computed the correlations between nearest neighbor (NN)
sites in the center leg of a three-leg ladder ($S_3$), and
between sites at the maximum distance along this leg ($S_4$).
$S_5$ and $S_6$ are the correlations between NN and next NN (NNN)
sites along the stripes.
The spin-spin correlations have been normalized in such a way that
their maximum (minimum) value is $+1$ ($-1$) for the z-components
of the two spins fully aligned or ferromagnetic (FM) (respectively
anti-aligned or AF). In Fig.~\ref{fig3}(a), corresponding to
$J/t=0.35$, $e_s=1.0$ it can be seen that the correlations
$S_1$ and $S_2$ across the stripe are positive at high
temperature and they increase as $T$ is lowered.
Around $T\approx 0.12$ these correlations reach their
maximum value and as the temperature is further lowered they
suddenly become fully AF and remain negative down to the lowest
temperature reached. These changes from FM to AF
indicate that the intervening mostly hole-free ladders are
in-phase (anti-phase) above
(below) $T\approx 0.11$. $S_3$ and $S_4$ (only shown in
Fig.~\ref{fig3}(a)) show a monotonic behavior as the temperature
is decreased. They indicate a full polarization of the spin domains
at low temperatures. A
similar behavior can be observed for smaller $e_s$ and
larger $J$ (Fig.~\ref{fig3}(b)). For the $e_s=1.0$ but $J=0.7$
(Fig.~\ref{fig3}(c)), $T^\ast_2$ increases and
it increases further for smaller density
($x=0.083$ on the $12\times 12$ cluster). The same features
survive when a $XY$ term in the exchange interaction is included,
as it can be seen in Fig.~\ref{fig3}(d) for $\gamma=0.25$
and 0.5. As the Heisenberg interaction is made more
isotropic, $T^\ast_2$ slightly decreases. Results for $\gamma =0.5$
suggest that the spin ladders are not going to be fully polarized
in the isotropic limit $\gamma =1.0$.\cite{kim}
As discussed below, the correlations along the stripe
$S_5$ and $S_6$ are much smaller than the previous ones and
hence they show a more erratic behavior, specially below the
crossover temperature. Above this temperature, these correlations
are AF, and the NN correlation is in general larger (in absolute
value) than the NNN one, although  for $J=0.35$ and just above
$T^\ast_2$ the opposite behavior is true in agreement with the
analysis made in Ref.~\onlinecite{martins} which is valid for small
$J/t$. A note of caution should be make. Since translational
invariance in the direction perpendicular to the stripes is broken,
the spin-spin correlations depends on the hole density on each site
(e.g. $\langle S^{z}_i S^{z}_i\rangle= \langle n_i\rangle$) which in
turn has a smooth variation with
temperature as shown in Fig.~\ref{fig1}(b)). However, the changes in
the spin-spin correlations, specially at low temperatures, are much
stronger than the variation of the hole density, so one could safely
ignore that dependency.

\begin{figure}
\begin{center}
\epsfig{file=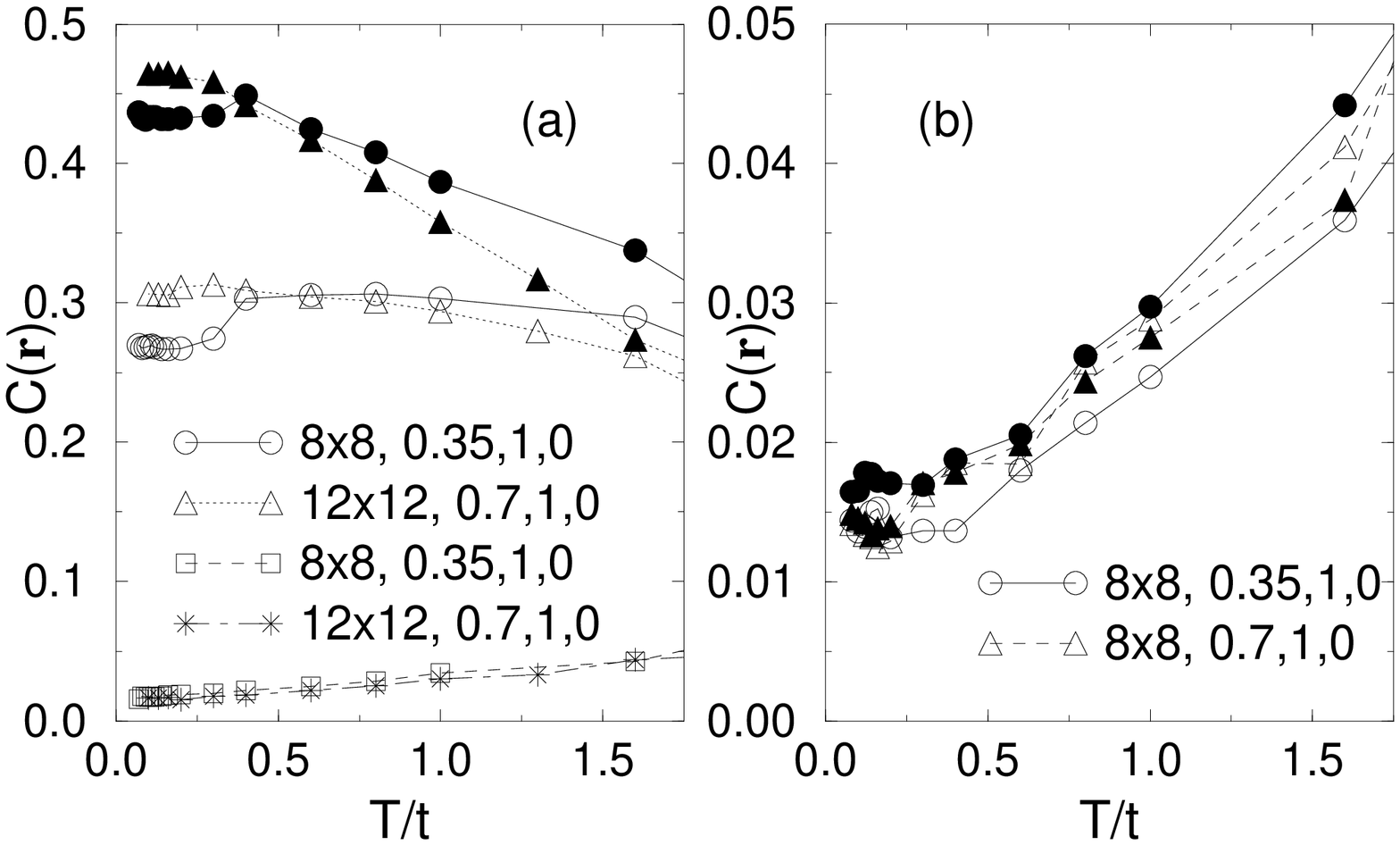,width=5cm,angle=-90}
\epsfig{file=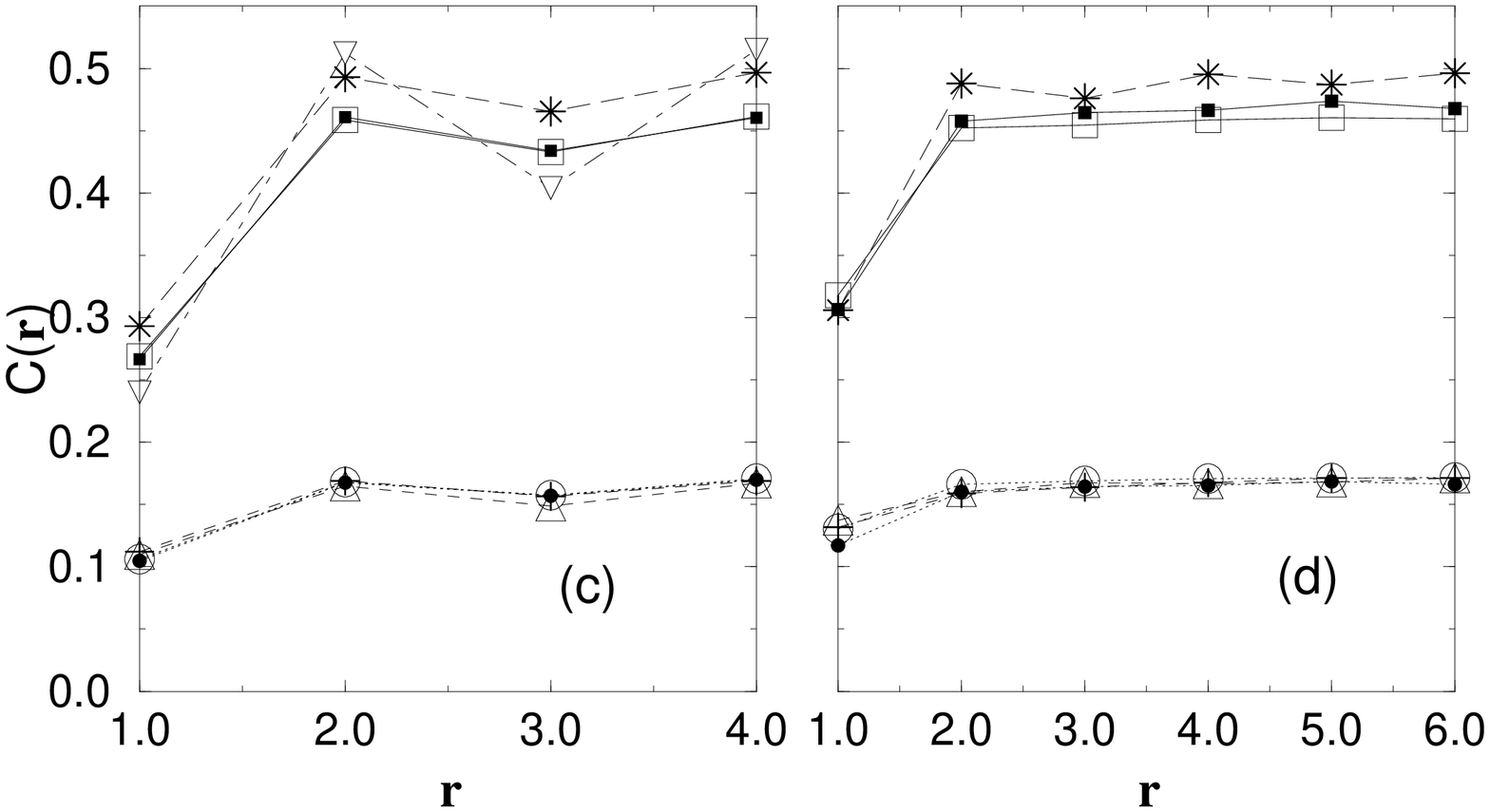,width=4.8cm,angle=-90}
\end{center}
\caption{(a) Hole-hole correlations along the stripe, $C_{s}$,  vs.
temperature. Lattice size, $J$, $e_s$, and $\gamma$ are indicated
on the plot. Open (full) circles and triangles correspond to NN
(maximum distance) sites. Squares and stars indicate $C_{s1}$ (see
text) at the maximum distance.
(b) Hole-hole correlations on the ``spin" ladder, $C_{c1}$.
Open (full) symbols correspond to NNN (maximum distance) sites.
(c) Hole-hole correlations for the $8\times 8$ cluster, $J=0.35$,
$e_s=1.0$, $\gamma=0$ above (open circles, squares and triangles) and
below (full circles and squares, plus) $T^\ast_2$.
$C_{s}$ (squares), $C_{s1}$ (circles), and $C_{1}$ (triangles, plus)
are shown. The last two correlations have been multiplied by 10.
The down triangles correspond to $C_{s}$ but for the metastable
in-phase state.
The stars indicate exact results for an eight site
$t-Jz$ ring at quarter filling at $T=T^\ast_2$.
(d) same as (c) but for the $12\times 12$ cluster, $J=0.7$,
$e_s=1.0$, and $\gamma=0$.
}
\label{fig4}
\end{figure}

Our final study concerns the other important issue that is the
relationship between stripes and hole pairing. A sign of hole
attraction is the presence of largest hole-hole correlations
at smallest distances. Taking into account the remark made
earlier about the broken translational invariance the hole-hole
correlations $C({\bf r}_i,{\bf r}_j) = \langle n_i n_j\rangle$
result proportional to the hole density at each site.
It is then expected that $C({\bf r}_i,{\bf r}_j)$ along the stripe,
$C_{s}({\bf r})$, as shown in
Fig.~\ref{fig4}(a), present a smooth increase as the temperature
is reduced, while the correlations along the first column next to
the stripe, $C_{1}$, are smoothly decreasing. However, near and
below $T^\ast_2$ these correlations behave roughly
independently of $T$. The same behavior can be observed for
$C({\bf r}_i,{\bf r}_j)$, with ${\bf r}_i$ on the center leg of a
three-leg ladder and ${\bf r}_j$ on the column next to it, $C_{c1}$,
(Fig.~\ref{fig4}(b)). The normalization adopted is such that:
\begin{eqnarray}
\sum_{y} C((x,y_0),(x,y)) = \langle N_{h,x}\rangle
\nonumber
\end{eqnarray}
\noindent
where $y_0$ is the $y$-coordinate of a reference site on column $x$,
and $\langle N_{h,x}\rangle$ is the number of holes on that column.
Then, $C({\bf r},{\bf r})=1$.

In Figs.~\ref{fig4}(c) and (d) we show $C({\bf r}_i,{\bf r}_j)$ at 
several points $({\bf r}_i,{\bf r}_j)$ after being averaged in a
region $\Delta T\approx 0.2$ above and below the crossover 
temperature. It may be noticed that there is no abrupt changes
as $T^\ast_2$ is crossed since the averaged correlations immediately
above and below it fall within each other error bars.
The first place to look for hole pairs are on the stripes, where
the largest hole density is located. The results for the hole-hole
correlations along the stripes (Fig.~\ref{fig4}(c)), for all the
parameter sets studied, show that $C_s(r)$ are 
{\em smallest} at nearest neighbor (NN) sites
and largest at the maximum possible distance, although they are
approximately constant beyond NN sites. This behavior is consistent
with a {\em metallic} behavior of the stripes, as expected in the
cuprates.\cite{ichikawa} Results for $J=0.7$ are virtually
indistinguishable from the ones for $J=0.35$, for the same 
$e_s=1$. In the metastable in-phase state, degenerate with the
anti-phase state within error bars at the same low temperatures,
the largest
correlations also occur at the largest distance but with a
more pronounced $\pi$-modulation, indicating a stronger
coupling with the spin surrounding.
The same short distance repulsion is obtained for $C_1(r)$ and also
when ${\bf r}_i$ belongs to the stripe and ${\bf r}_j$ to the first
column next to it ($C_{s1}$) (Fig.~\ref{fig4}(c)).
In Fig.~\ref{fig4}(d), similar results are shown for the $12\times 12$
cluster, $x=0.083$. An alternative scenario\cite{castroneto} assumes
that hole pairs with $d_{x^2-y^2}$ symmetry
are formed due to short-range AF correlations
inside the spin domains located in
between the stripes as well-known results on small cluster
calculations show.\cite{rierayoung} Again, due to the dependence of
the hole-hole correlations with the local hole density, the
correlations between sites along the central leg of the spin domains
are extremely small and hence they are almost completely masked by
errors. However, the correlations between a site on the central leg
of the three-leg ladder and a site on the first column next to it,
which are almost identically zero on the $12\times 12$ cluster,
acquire on the $8\times 8$ cluster and $x=0.125$, very similar values
to those of $C_{s1}$. These correlations could then become important
as the hole density is further increased. Fig.~\ref{fig4}(c),(d) also
show exact diagonalization results for hole-hole correlations on
$L=8$ and $L=12$ $t-Jz$ rings at quarter filling ($T=T^\ast_2=0.08$,
0.18 respectively). They are very similar to the correlations
$C_{s}$ along the stripes except for a small shift due to the
fact that the hole density on stripes is slightly smaller than
$x=0.5$. Besides, as shown earlier, $S_1$ shows that contiguous
spin ladders are almost completely AF coupled in a 2D square
lattice. This combined behavior suggests, at least for the $t-Jz$
limit, a generalized spin-charge separation, or more properly, a
separation between the spin background and the
stripes.\cite{chernyshev,martins}

In summary, we have studied an anisotropic $t-J$ model, close to
the Ising limit, where straight site-centered stripes are imposed
by an on-site potential reflecting a mechanism which is not intrinsic
to the 2D short-range electronic correlations of that model.
The results of the present study suggest that we are able to study
with QMC simulations
the temperature region between the formation of the stripes at a charge
ordering temperature (more or less arbitrarily defined in our model)
and the spin ordering process which takes place at a much lower
temperature. This lower crossover, at which the spin domains become
anti-phase domains and an incommensurate magnetic order appears, should 
correspond in the cuprates to the opening of the pseudogap.
In this sense, this result that stems from a model where the stripes
are caused by long-range Coulomb interaction, electron-phonon
couplings, or other mechanisms that are described by an on-site
potential, is at variance with recent results in which the stripes
are originated in the pure $t-J$ model.\cite{white,martins}
In that and other approaches\cite{zaanen}, the stripes are the
{\em consequence} of anti-phase domain formation and both features
should occur simultaneously. The results of the present study,
including the behavior of spin-spin correlations along the stripes,
open the possibility of experimentally discriminate the mechanism
leading to stripe formation and hence to determine to what extent the
stripes are universal to the cuprates or depend on particular details
of the various compounds. The hole-hole correlations along and near
the stripes show a metallic behavior with no indications of hole
attraction. The question arises if a hole attraction on the stripes
could appear by taking the isotropic Heisenberg term in the model.
One should take into account that, as previous exact results 
show,\cite{tjz} hole attraction is actually enhanced in the $t-Jz$
model with respect to the fully isotropic $t-J$ model. Besides, to
give more support to our result, it has been suggested that stripes
could be introduced in a uniform $t-J$ model by taking an Ising
spin interaction at the stripes links.\cite{eroles}
An improvement on statistical errors in order to deal with larger
hole densities and lower temperatures would be necessary to detect
signs of hole attraction inside the intervening regions between
stripes.\cite{Note}

\acknowledgements

We wish to acknowledge many interesting discussions with A. Castro Neto,
C. Gazza, G. B. Martins, and A. Trumper.
We thank the Supercomputer Computations Research Institute (SCRI)
and the Academic Computing and Network Services at Tallahassee
(Florida) for allowing us to use their computing facilities.


\begin{references}
\bibitem{Note} Many of the features observed below $T^\ast_2$, assuming
     the metallic behavior remains down to zero temperature, 
     can be reproduced by a ``toy model" consisting of a $2 \times 8$
     or $2 \times 12$ ladders with the same Hamiltonian as
     defined by Eqs.~(\ref{ham_anis})-(\ref{stripepot}), and two holes.
     This ``toy model" would allow the computation of dynamical
     correlations.\cite{monte2}
\end{references}
\end{document}